\documentclass[final,3p,times,twocolumn]{elsarticle}

\usepackage{amsmath,amssymb,amsfonts}
\usepackage{textcomp}
\usepackage{comment}
\usepackage{subcaption}
\usepackage{gensymb}

\begin{document}

\begin{frontmatter}

\title{Tumor monitoring and detection of lymph node metastasis using quantitative ultrasound and immune cytokine profiling in dogs undergoing radiation therapy: a pilot study}

\author{Mick Gardner$^{a,b}$}
\author{Audrey Billhymer$^{c}$} 
\author{Rebecca Kamerer$^{c}$}
\author{Joanna Schmit$^{c}$}
\author{Trevor Park$^{d}$}
\author{Julie Nguyen-Edquilang$^{c}$}
\author{Rita Miller$^{b}$}
\author{Kim A Selting$^{c}$}
\author{Michael Oelze$^{b}$}

\affiliation{organization={Corresponding Author: Beckman Institute, University of Illinois Urbana-Champaign},
            addressline={405 N Matthews Ave}, 
            city={Urbana}, 
            state={IL},
            postcode={61801},
            country={USA}}

\affiliation{organization={The Grainger College of Engineering, Department of Electrical and Computer Engineering, University of Illinois at Urbana-Champaign},
            addressline={306 N Wright St}, 
            city={Urbana}, 
            state={IL},
            postcode={61801},
            country={USA}}

\affiliation{organization={Department of Veterinary Clinical Medicine, University of Illinois Urbana-Champaign},
            addressline={306 N Wright St}, 
            city={Urbana}, 
            state={IL},
            postcode={61801},
            country={USA}}

\affiliation{organization={Department of Statistics, University of Illinois Urbana-Champaign},
            addressline={306 N Wright St}, 
            city={Urbana}, 
            state={IL},
            postcode={61801},
            country={USA}}

\begin{abstract}
Quantitative ultrasound (QUS) characterizes the composition of cells to distinguish diseased from healthy tissue. QUS can reflect the complexity of the tumor and detect early lymph node (LN) metastasis ex vivo. The objective in this study was to gather preliminary QUS and cytokine data from dogs undergoing radiation therapy and correlate QUS data with both LN metastasis and tumor response. Spontaneous solid tumors were evaluated with QUS before and up to one year after receiving RT. Additionally, regional LNs were evaluated with QUS in vivo, then excised and examined with histopathology to detect metastasis. Paired t-tests were used to compare QUS data of metastatic and non-metastatic LNs within patients. Furthermore, paired t-tests compared pre- versus post-RT QUS data. Serum was collected at each time point for cytokine profiles. Most statistical tests were underpowered to produce significant \textit{p} values, but interesting trends were observed. The lowest \textit{p} values for LN tests were found with the envelope statistics K (\textit{p} = 0.142) and \(\mu\) (\textit{p} = 0.181), which correspond to cell structure and number of scatterers. For tumor response, the lowest \textit{p} values were found with K (\textit{p} = 0.115) and \(\mu\) (\textit{p} = 0.127) when comparing baseline QUS data with QUS data 1 week after RT. Monocyte chemoattractant protein 1 (MCP-1) was significantly higher in dogs with cancer when compared to healthy controls (\textit{p} = 1.12e-4). A weak correlation was found between effective scatterer diameter (ESD) and Transforming growth factor beta 1 (TGF\(\beta\)-1), with a Pearson’s correlation coefficient of -0.35. While statistical tests on the preliminary QUS data alone were underpowered to detect significant differences among groups, our methods create a basis for future studies. 
\end{abstract}


\end{frontmatter}

\section{Introduction}
Cure rates for cancer are high when local disease can be controlled with surgery, radiation, or both, so long as metastasis does not occur. With established metastasis, control is challenging and cure is rare. Untreated or resistant tumors can cause significant pain, dysfunction, and decreased quality of life. Furthermore, successful response of tumors to radiation therapy (RT) can include minimal or lack of tumor shrinkage, making it difficult to predict which tumors will progress over time. This creates a necessity for other methods to detect or quantify the tumor’s response to treatment, motivating possible changes in the treatment plan to improve prognosis.

Quantitative ultrasound (QUS) is a novel way to use sound waves to characterize the composition of different tissues \cite{oelze_review_2016}. Compared to other imaging modalities, ultrasound is attractive due to its low cost, portability, and lack of ionizing radiation. In addition to traditional B-mode images, which simply display the amplitude of received echoes, additional parameters can be calculated from the raw radio-frequency (RF) data of an ultrasound signal which allows QUS analysis. Such parameters include spectral-based parameters, which are derived from the backscatter coefficient (BSC) of tissue and include effective scatterer diameter (ESD) and effective acoustic concentration (EAC) \cite{insana_identifying_1992}. Other parameters can be derived from the statistical distribution of RF envelope values. These include the K statistic, which is indicative of cell spacing and structure, and the \(\mu\) value, which corresponds to the number of scatterers per resolution cell \cite{jakeman_statistics_1980}.

Quantitative ultrasound has been applied to early detection of cell death in response to cancer treatments. In mouse models of head and neck cancer treated with radiation, QUS was able to detect response to radiation therapy as early as 24 hours after treatment before visible tumor shrinkage was observed \cite{vlad_quantitative_2008}. This was also found in human patients with head and neck tumors \cite{tran_quantitative_2020}. Another study found that QUS parameters also correlated with early in vivo tumor cell death in human breast cancer patients with locally advanced disease treated with chemotherapy during the first month of treatment when compared to patients who did not respond \cite{sadeghi-naini_quantitative_2013}. Additional studies in women with breast cancer found that QUS parameters were particularly useful in distinguishing responders from non-responders early after treatment by chemotherapy when combined with texture features as additional parameters for statistical testing and classification \cite{sannachi_response_2018,sannachi_breast_2019}. Lastly, the acoustic concentration (EAC) experienced major increases in response to cell death in prostate tumors in mice responding to ultrasound microbubble and hyperthermia treatment \cite{sharma_vivo_2021}.

In addition to tumor monitoring, QUS has been used to detect LN metastasis as well. Traditional approaches to screen LNs for metastasis include injecting a radioactive tracer, a dye, or both into or around a tumor, and then performing surgery to remove the sentinel LNs that receive those tracers as they drain from the tumor. Quantitative ultrasound techniques offer a non-invasive method of LN evaluation and have been successfully used to identify early LN metastasis. However, these studies were initially conducted on surgically excised LNs \cite{mamou_three-dimensional_2010,mamou_three-dimensional_2011}, and only recently on in situ LNs \cite{hoerig_classification_2023}. While QUS will not replace histologic assessment of a LN, QUS could be a novel way to determine sentinel LNs before surgery is performed.

This pilot study was done to identify trends in QUS data and attempt to correlate QUS findings with LN metastasis and tumor response in dogs undergoing stereotactic radiation therapy (SRT). We compared baseline QUS data between responsive and non-responsive patients, tracked QUS parameters in tumors for up to 1 year after radiation therapy, and compared QUS findings in LNs with and without metastasis. Because radiation therapy can impact the interaction between tumor cells and the immune system, we also collected serial serum samples to develop immune profiles. We hypothesized that there would be some statistical separation in the QUS data between responsive versus non-responsive tumors, and between metastatic versus non-metastatic LNs. We also hypothesized that there would be shifts in cytokine profiles correlated with shifts in QUS data that reflect radiation-induced immunogenic cell death.

\section{Methods}

\subsection{Pre-clinical trial}
\subsubsection{Patients}
Dogs of any age, breed, sex, and weight with solid tumors were eligible for inclusion. A macroscopic tumor with histologic or cytologic diagnosis was required. Tumors with a low risk of LN metastasis (such as bone or soft tissue sarcomas) were excluded. Client-owned dogs that were presented to the Oncology service at the University of Illinois Veterinary Teaching Hospital, Urbana, IL, were recruited and all procedures were approved by the Institutional Animal Care and Use Committee (IACUC) at the University of Illinois at Urbana-Champaign (ID \# 21124, June 1, 2021).

\begin{figure*}[!t]
    \centering
    \begin{subfigure}{0.32\linewidth}
        \includegraphics[scale=0.22]{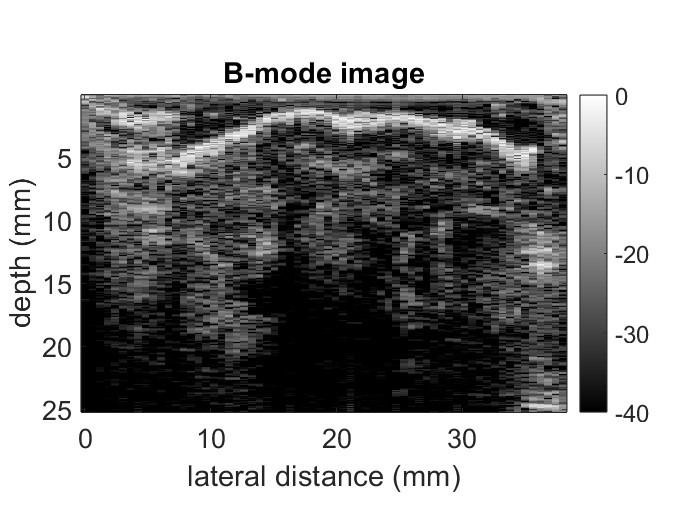}
        \caption{}
        \label{subfig1a:map_bmode}
    \end{subfigure}
    ~
    \begin{subfigure}{0.32\linewidth}
        \includegraphics[scale=0.19]{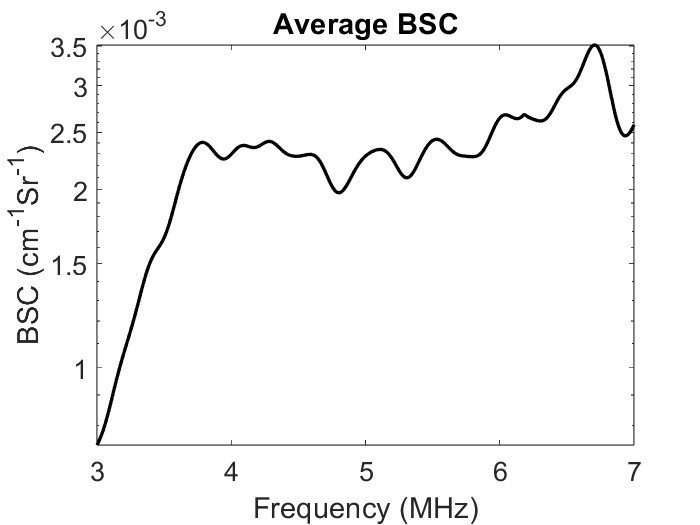}
        \caption{}
        \label{subfig1b:map_bsc}
    \end{subfigure}
    ~
    \begin{subfigure}{0.32\linewidth}
        \includegraphics[scale=0.22]{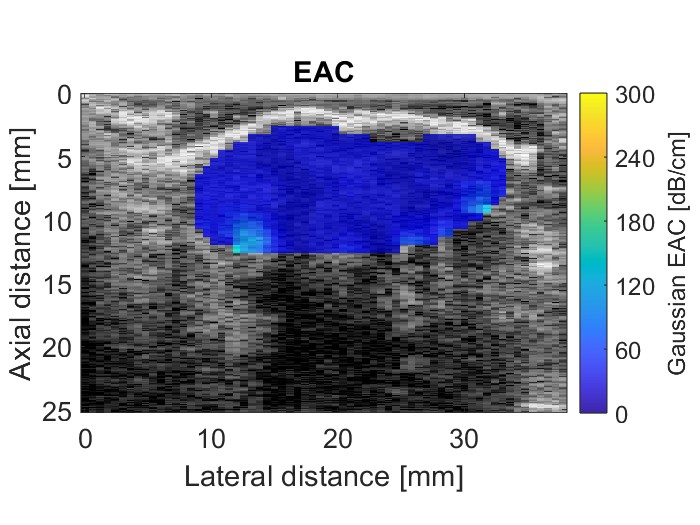}
        \caption{}
        \label{subfig1c:map_eac}
    \end{subfigure}\\
    \begin{subfigure}{0.32\linewidth}
        \includegraphics[scale=0.22]{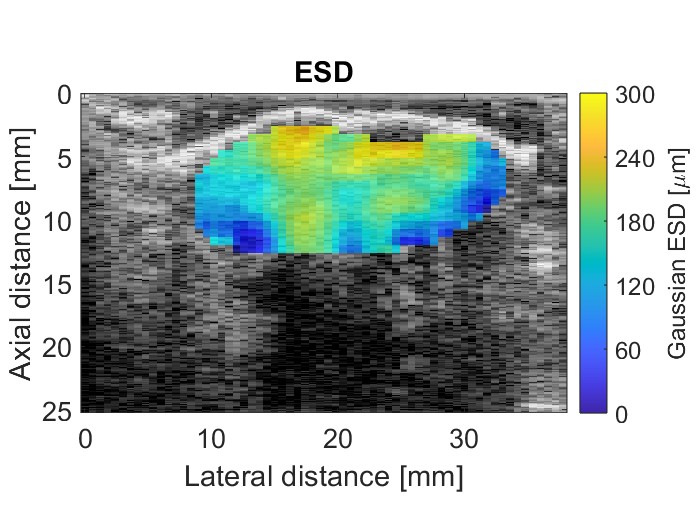}
        \caption{}
        \label{subfig1d:map_esd}
    \end{subfigure}
    ~
    \begin{subfigure}{0.32\linewidth}
        \includegraphics[scale=0.22]{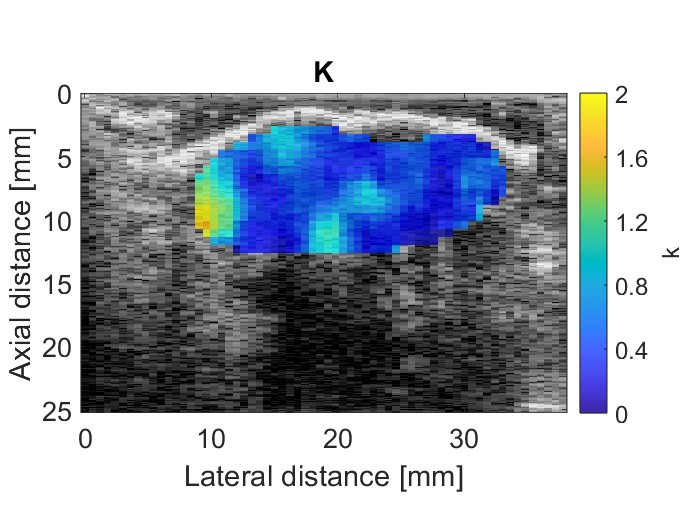}
        \caption{}
        \label{subfig1e:map_k}
    \end{subfigure}
    ~
    \begin{subfigure}{0.32\linewidth}
        \includegraphics[scale=0.22]{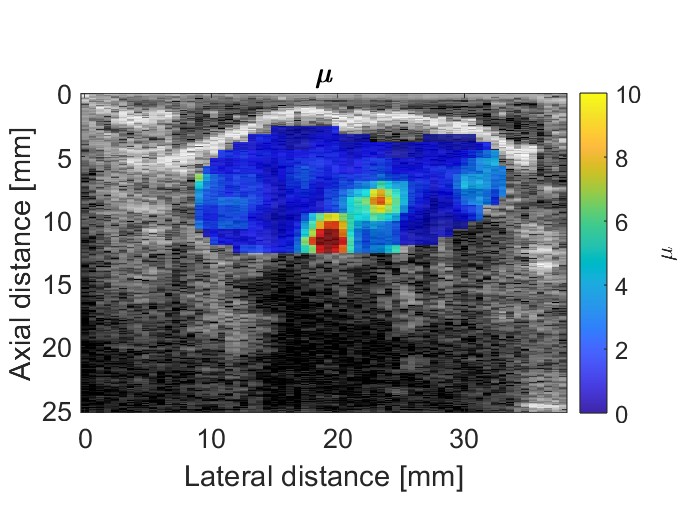}
        \caption{}
        \label{subfig1f:map_mu}
    \end{subfigure}\\
    \caption{Example QUS maps. (a) Example B-mode image with (b) its corresponding averaged BSC, and distributions of (c) EAC, (d) ESD, (e) K, and (f) \(\mu\).}
    \label{fig1:qus_maps}
\end{figure*}

\subsubsection{Procedures}
Dogs were evaluated with a minimum of complete blood count, serum chemistry profile, urinalysis, staging tests appropriate for the tumor type (thoracic and/or abdominal radiographs and/or ultrasound), and a computed tomography (CT) scan of the affected area for the dual purpose of LN assessment for staging and to generate images used for RT planning. Computed tomography findings were used to determine which regional LNs would be removed based on size and shape with the intent to excise any abnormal and at least one normal appearing LN. Dogs for which all regional LNs appeared abnormal were not eligible.

At baseline and each visit thereafter, the primary tumor was evaluated with QUS. Two to four regional LNs were identified for planned excision, evaluated with QUS prior to surgery, then submitted for histopathology after excision. Dogs were followed until progressive disease, death due to other causes, or for up to 1 year after radiation therapy, whichever came first. Ultrasound scans consisted of four images per time point for the tumors, and four images for each LN to be excised. Ultrasound scans of the tumor were performed before surgical LN removal, before the first dose of RT, and immediately after the last dose of RT, with follow up scans done at 2-week, 3-week, 4-week (1-month), 2-month, 3-month, 6-month, and 1-year marks from the start of the trial. Ultrasound scans were performed using an L14-5/38 probe connected to a SonixRP imaging device (BK Ultrasound, Peabody, MA, USA), transmitting at 10 MHz.

Radiation therapy was delivered using a Varian TrueBeam® v.2.7 linear accelerator. Treatment planning was performed using the Varian Eclipse® treatment planning system v.15 (Varian Medical Systems, Inc, Palo Alto, CA, USA). The Gross Tumor Volume (GTV) was contoured, and a 1-2 mm isometric expansion was added to create a Planning Target Volume (PTV). The radiation protocol consisted of three fractions of 8 Gy given every 24-48 hours over the course of 3-5 days, with the dose normalized to 100\% of the dose delivered to 95\% of the PTV. All treatment plans underwent quality assurance using SunNuclear PerFraction EPID-based dosimetry, with passing criteria defined as an overall gamma of greater than or equal to 97\% using a 3\% within 3 mm setting for each point evaluated.

\subsection{Quantitative Ultrasound (QUS) Parameters}
Four QUS parameters were used for statistical evaluation, two of which are based on the backscatter coefficient (BSC), while the other two are based on the statistics of radio-frequency (RF) envelope values. The BSC of a tumor or LN was estimated using the reference phantom method, \cite{oelze_review_2016}. A commercial uniform phantom (CIRS, Inc., Norfolk, VA, USA, part no. 14090501) was used as the reference. This phantom was scanned during the first month of the trial. The imaging settings used to scan the reference were saved to a preset on the SonixRP system, and that preset was loaded for each scan of each patient throughout the trial. This ensured that all scans were conducted with identical settings, and allowed the same reference to be used for QUS analysis of each patient. The BSC was estimated over the bandwidth of 3 MHz to 7 MHz. From the BSC, the effective scatterer diameter (ESD) and the effective acoustic concentration (EAC) were estimated by fitting a spherical Gaussian form factor to the estimated BSC \cite{oelze_characterization_2002}. These parameters roughly correspond to cell size and scattering strength respectively. 

Two envelope statistics were also estimated. The K and \(\mu\) values from a homodyned-K distribution were estimated by fitting its PDF to the histogram of envelope values \cite{mamou_three-dimensional_2011}. The parameter \(\mu\) corresponds to the number of scatterers per resolution cell, while the parameter K corresponds to scatterer periodicity, in that a higher K implies a more regular cell spacing and structure than a lower K \cite{dutt_speckle_1995}.

\begin{table*}[t]
    \centering
    \begin{tabular}{|c|c|c|c|c|c|}
    \hline
         Patient Number&  Sex&  Breed&  Volume Response (\%)&  LN Metastasis&  Notes\\
         \hline
         1&  MC&  Mix&  -94.5*&  N/A&  Lost LN Data\\
         2&  FS&  Pomeranian&  -100*&  None&  Died at 1 month\\
         3&  FS&  Mix&  -2.76&  None&  Progressive at 4 months\\
         4&  FS&  AmStaff&  -64.6*&  N/A&  Could not find LNs\\
         5&  FS&  Boxer&  -80.8*&  None&  Off-trial at 3 months\\
         6&  FS&  Scottish Terrier&  -80.0*&  Left Mandibular&  \\
         7&  FS&  Shih Tzu&  -87.3*&  None&  \\
         8&  FS&  Mix&  -62.3*&  None&  Off-trial at 2 months\\
         9&  MC&  Miniature Schnauzer&  -88.2*&  None&  \\
         10&  MC&  Mix&  -100*&  Right Mandibular&  \\
         11**&  MC&  &  N/A&  Right Mandibular&  \\
         \hline
    \end{tabular}
    \caption{Patient information. Those that were taken off trial early are noted in the “Notes” column. All other patients were on trial for the full year. MC – Male Castrated. FS – Female Spayed. *Marks classification as a responder based on WHO metrics. **Patient 11 added for LN analysis only.}
    \label{tab:patient_info}
\end{table*}

\begin{table*}[t]
    \centering
    \begin{tabular}{|c|c|c|c|c|}
        \hline
         Parameter&  Non-metastatic&  Metastatic&  \textit{p} value paired&  \textit{p} value unpaired\\
         \hline
         K&  0.515 \(\pm\) 0.143&  0.504 \(\pm\) 0.0209&  0.142&  0.661\\
         \(\mu\)& 2.50 \(\pm\) 0.985&  2.32 \(\pm\) 0.333&  0.181&  0.238\\
         ESD&  118 \(\pm\) 23.7&  131 \(\pm\) 26.6&  0.743&  0.638\\
         EAC&  39.0 \(\pm\) 6.79&  32.5 \(\pm\) 7.52&  0.598&  0.372\\
         HT2&  N/A&  N/A&  0.509*&  0.466\\
         \hline
    \end{tabular}
    \caption{Lymph node data. The first two columns show mean \(\pm\) std for each QUS parameter. The last two columns show \textit{p} values for t-tests. HT2 = Hotelling’s T-squared. *Test on K, \(\mu\) only because there were too few data points to include all four parameters.}
    \label{tab:ln_data}
\end{table*}

Each of these parameters was calculated inside a hand-drawn region of the ultrasound images which segmented the tumor or LN from surrounding tissue. After the segmentation, regions of interest were defined within the segment with a width and height of 2 mm x 2 mm, overlapping by 75\%, and QUS values were estimated in these ROIs. This created distributions of values for each QUS parameter over the tumor/LN such as those shown in Figure \ref{fig1:qus_maps}. To create data points, the mean of each of these regions was taken. Then, since each tumor/LN was scanned four times each session, the means from same-day/same-node scans were averaged together to create the independent data points which were used for statistical tests.

\subsection{Statistical Tests}
\subsubsection{LN Classification}
Two types of LN classification tests were performed on the data. The first set consisted of paired tests using one metastatic and one non-metastatic LN from within each patient that had a metastatic LN. Paired t-tests were done for each QUS parameter, as well as a paired Hotelling’s T-squared test using all four QUS parameters. The other set of tests consisted of independent t-tests using one LN from each patient. The metastatic LN was chosen from those patients which had metastasis, while one non-metastatic LN was chosen from patients which did not show metastasis. Similarly to before, unpaired t-tests were done for each QUS parameter, as well as a Hotelling’s T-squared test on all four parameters. For the t-tests, Bonferroni correction was applied for multiple comparisons. An overall significance level of 0.05 was desired, so with four comparisons, a significance level of 0.05/4 = 0.0125 was used to determine significance for individual tests. The Hotelling’s T-squared tests simply used a significance level of 0.05.

\begin{figure}[!t]
    \centering
    \begin{subfigure}{0.475\columnwidth}
        \includegraphics[scale=0.16]{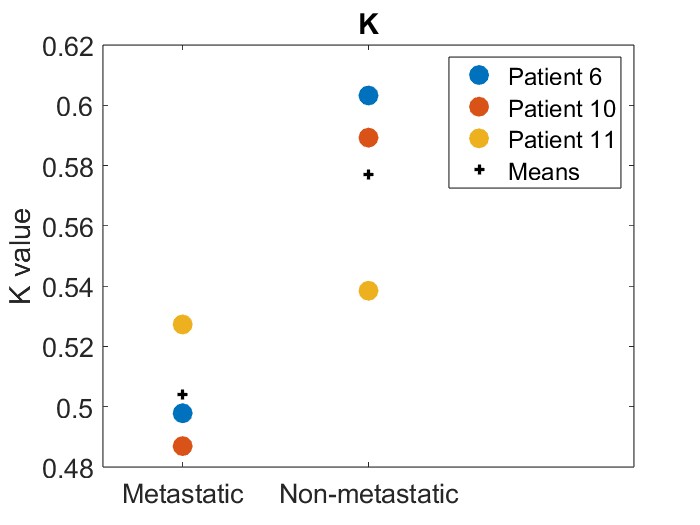}
        \caption{}
        \label{subfig2a:ln_paired_k}
    \end{subfigure}
    ~
    \begin{subfigure}{0.475\columnwidth}
        \includegraphics[scale=0.16]{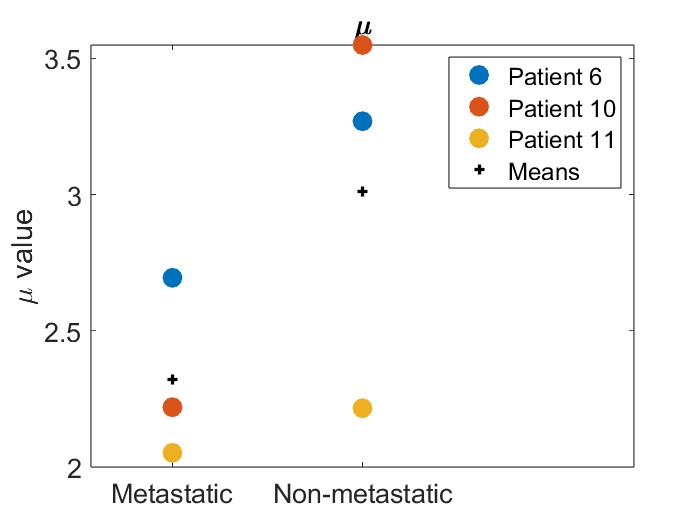}
        \caption{}
        \label{subfig2b:ln_paired_mu}
    \end{subfigure}\\
    \begin{subfigure}{0.475\columnwidth}
        \includegraphics[scale=0.16]{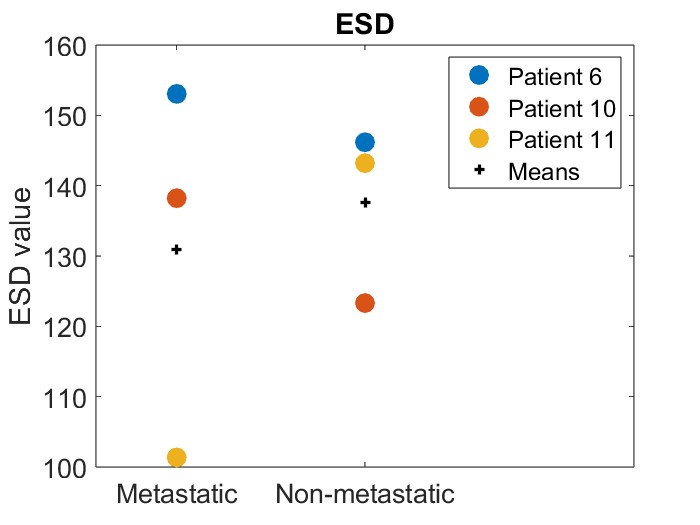}
        \caption{}
        \label{subfig2c:ln_paired_esd}
    \end{subfigure}
    ~
    \begin{subfigure}{0.475\columnwidth}
        \includegraphics[scale=0.16]{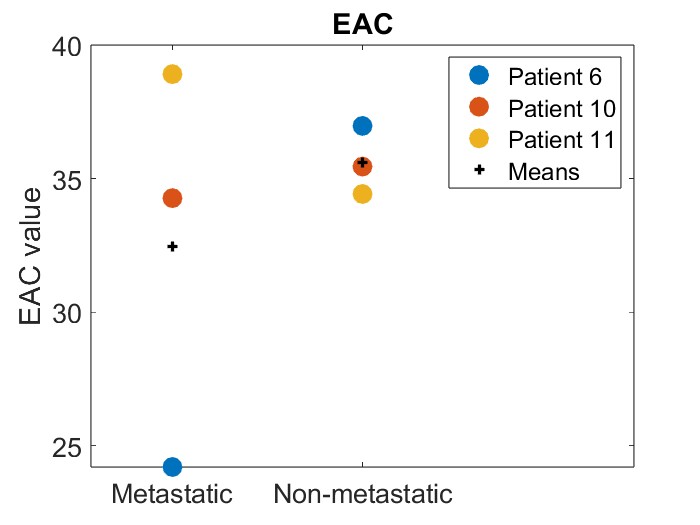}
        \caption{}
        \label{subfig2d:ln_paired_eac}
    \end{subfigure}
    \caption{Data points for paired LN t-tests. This dataset contains one metastatic and one non-metastatic LN from patients 6, 10, and 11. (a) K, (b) \(\mu\), (c) ESD, (d) EAC.}
    \label{fig2:ln_paired_points}
\end{figure}

\subsubsection{Tumor response detection}
This set of tests compared baseline QUS data to post-RT data in patients which responded to RT. The baseline data was the time point prior to SRT but after surgical removal of LNs. Four paired t-tests for each QUS parameter and four paired Hotelling’s T-squared tests were done to compare these baseline points with individual time points after RT. These time points were the 1-week, 2-week, 3-week and 4-week (1 month) marks of the trial. The same significance levels as described in section 2.3.1 were used to determine significance.

\subsubsection{QUS and Cytokine Correlation}
This analysis compared QUS data with serum cytokine data collected throughout the trial using customized Multiplex assays (Millipore Sigma, EMD Millipore Corporation, Burlington, MA, USA). The specific cytokines measured were Interferon gamma (IFN-\(\gamma\)), Interleukin-2 (IL-2), Interleukin-6 (IL-6), Interleukin-8 (IL-8), Interleukin-10 (IL-10), Interleukin-15 (IL-15), Interleukin-18 (IL-18), Interferon gamma-inducible protein (IP-10), Monocyte chemoattractant protein-1 (MCP-1), Transforming growth factor beta 1 (TGF\(\beta\)-1), TGF\(\beta\)-2, and TGF\(\beta\)-3. Cytokine data was collected at all the same time points as QUS data for each patient on the trial. Additionally, one data point was taken for each cytokine on each of 10 healthy patients as a control group. Student’s t-tests compared baseline (pre-RT) data from cancer patients to data from control patients. With 13 cytokines, Bonferroni correction was applied to the standard significance value of \(\alpha\) = 0.05, producing an overall significance level of \(\alpha\)/13 = 0.05 / 13 = 0.0042, which was used to determine statistical significance for these tests. Lastly, correlation coefficients were estimated between each QUS parameter and each cytokine.

\section{Results and Discussion} \label{Results_and_discussion}
\subsection{Pre-clinical Trial}
Of the 10 patients that were enrolled, 9 of them had a volume response of at least 50\% shrinkage, meaning they were classified as responders according to World Health Organization (WHO) metrics (Table \ref{tab:patient_info}). Only two of the original 10 patients showed metastasis in at least one of their excised LNs. Therefore, an eleventh patient was enrolled in the trial and received ultrasound scans of LNs. Patient 11 met all the criteria for enrollment but did not receive SRT or ultrasound monitoring of the tumor and was only used for LN analysis. For patient 1, LN data files were corrupted and lost. For patient 4, neither ultrasound nor surgery could locate the LNs. The rest of the patients did not show metastasis in any LNs in histopathology. In total, 24 LNs were scanned, with 3 of them being metastatic while the other 21 were not.

\begin{figure}[!t]
    \centering
    \begin{subfigure}{0.475\columnwidth}
        \includegraphics[scale=0.16]{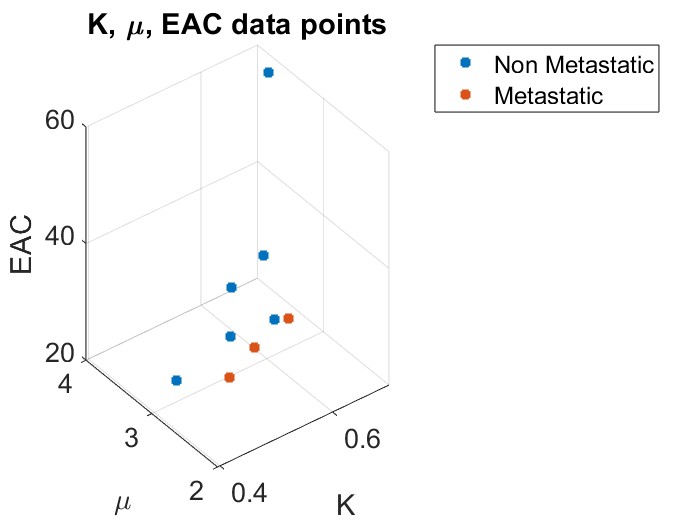}
        \caption{}
        \label{subfig3a:ln_unpaired_k_mu_eac}
    \end{subfigure}
    ~
    \begin{subfigure}{0.475\columnwidth}
        \includegraphics[scale=0.16]{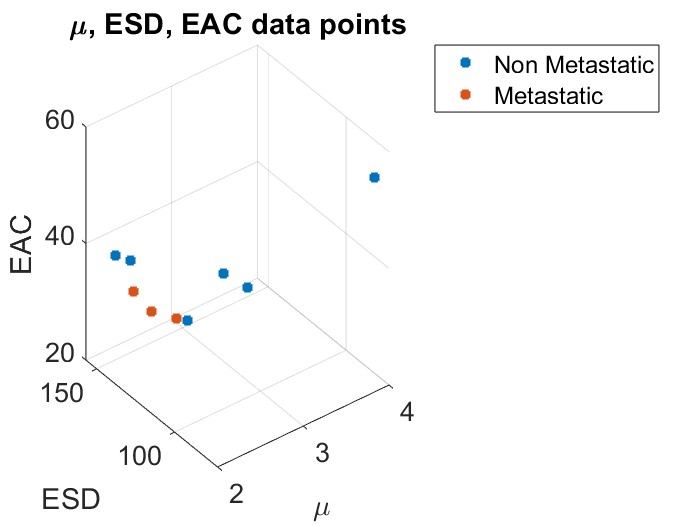}
        \caption{}
        \label{subfig3b:ln_unpaired_mu_esd_eac}
    \end{subfigure}\\
    \caption{Data points for unpaired LN t-tests. The metastatic LNs from patients 6, 10, and 11 were compared against one non-metastatic LN each from the other patients.}
    \label{fig3:ln_unpaired_points}
\end{figure}

\subsection{Lymph Node Classification}
Both K and \(\mu\) showed separation between metastatic and non-metastatic lymph nodes within the same patient (Figure \ref{fig2:ln_paired_points}). This separation resulted in K and \(\mu\) having much lower \textit{p} values in paired t-tests than either ESD or EAC (Table \ref{tab:ln_data}). There was also slight separation observed in the QUS data when comparing metastatic and non-metastatic lymph nodes across all patients (Figure \ref{fig3:ln_unpaired_points}). In this comparison, \(\mu\) and EAC had the lowest \textit{p} values (Table \ref{tab:ln_data}). However, no tests resulted in statistically significant \textit{p} values, likely due to a small sample size and large within-group variance.

\begin{table*}[t]
    \centering
    \begin{tabular}{|c|c|c|c|c|c|}
        \hline
         Parameter&  Baseline&  Week 1&  Week 2&  Week 3&  Week 4\\
         \hline
         K&  0.544 \(\pm\) 0.0750&  0.551 \(\pm\) 0.0666&  0.493 \(\pm\) 0.0776&  0.527 \(\pm\) 0.0779&  0.524 \(\pm\) 0.0636\\
         \(\mu\)&  3.06 \(\pm\) 0.970&  3.03 \(\pm\) 0.821&  2.43 \(\pm\) 0.878&  2.66 \(\pm\) 0.805&  2.62 \(\pm\) 0.726\\
         ESD&  112 \(\pm\) 28.2&  109 \(\pm\) 39.6&  117 \(\pm\) 36.4&  119 \(\pm\) 32.1&  119 \(\pm\) 26.9\\
         EAC&  34.6 \(\pm\) 10.3&  38.4 \(\pm\) 12.9&  35.9 \(\pm\) 16.3&  35.1 \(\pm\) 12.9&  34.1 \(\pm\) 9.55\\
         \hline
    \end{tabular}
    \caption{Mean and standard deviation for each QUS parameter from baseline (before RT) up to week 4 of the trial. Week 1 of the trial corresponds to immediately after the last dose of RT. Week 4 of the trial corresponds to 3 weeks after RT finished. }
    \label{tab:tumor_mean_std}
\end{table*}

These tests were performed to see whether there was a statistically significant difference in the QUS data between metastatic and non-metastatic LNs in our preliminary data. An important feature of our study design was its paired nature, in that metastatic and non-metastatic LNs from within the same patient were compared against each other, in addition to comparing LNs across multiple patients as other studies have done \cite{mamou_three-dimensional_2010,mamou_three-dimensional_2011,hoerig_classification_2023}. This allowed the use of paired tests which removed subject effects and ensured that differences correlate to metastasis, rather than some other biological difference such as breed or sex. While the statistical tests did not result in \textit{p} values less than 0.05, some separation was still observed in the data itself. This separation suggests that at least the K and \(\mu\) parameters could be reflecting metastasis despite major subject variance. With a larger sample size, perhaps more significant separation can be found.

\begin{figure}[!t]
    \centering
    \begin{subfigure}{\columnwidth}
        \includegraphics[scale=0.3]{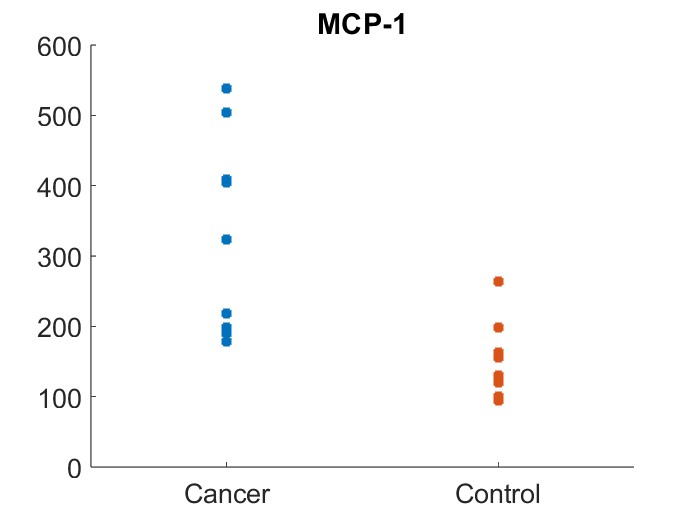}
        \caption{}
        \label{subfig4a:cytokine_mcp1_cancer_control}
    \end{subfigure}
    \caption{MCP-1 Cancer patients versus Control patients data points.}
    \label{fig4:cytokine_cancer_vs_control}
\end{figure}

\subsection{Tumor Response Detection}
The lowest \textit{p} values achieved were K (\textit{p} = 0.115) and \(\mu\) (\textit{p} = 0.127) when comparing baseline tumor QUS data with week two data. However, none of these tests resulted in significant \textit{p} values (Tables \ref{tab:tumor_mean_std} \& \ref{tab:tumor_p_value}). These tests were performed to determine whether the QUS parameters changed in a predictable way for patients that responded to SRT. Similar to the LN statistical tests, the paired design of our study accounts for inter-patient variability. Furthermore, while these results were also underpowered to produce statistically significant \textit{p} values, the lower \textit{p} values in K and \(\mu\) could suggest that these parameters are responding somewhat to changes in the tumor as a result of radiation. Again, a larger sample size would be needed to verify these results.

\begin{table}[t]
    \centering
    \begin{tabular}{|c|c|c|c|c|}
        \hline
         Parameter&  Week 1&  Week 2&  Week 3&  Week 4\\
         \hline
         K&  0.838&  0.115&  0.254&  0.453\\
         \(\mu\)&  0.952&  0.127&  0.243&  0.201\\
         ESD&  0.681&  0.447&  0.475&  0.253\\
         EAC&  0.216&  0.680&  0.882&  0.862\\
         HT2&  0.581&  0.625&  0.425&  0.422\\
         \hline
    \end{tabular}
    \caption{\textit{p} values for every paired statistical test of tumor monitoring. These \textit{p} values are the result of comparing baseline data (pre-RT) with corresponding weeks of the trial after RT. Week 1 corresponds to the time point immediately after RT finished. Week 4 corresponds to 3 weeks after RT finished. HT2 = Hotelling’s T-squared}
    \label{tab:tumor_p_value}
\end{table}

\begin{table}[t]
    \centering
    \begin{tabular}{|c|c|}
        \hline
         Cytokine&  \textit{p} value\\
         \hline
         IFNG&  0.88\\
         IL-6&  0.48\\
         IL-8&  0.30\\
         IL-15&  0.98\\
         IP-10&  0.48\\
         IL-18&  0.29\\
         MCP-1&  1.12E-4\\
         TGF\(\beta\)-1&  0.81\\
         TGF\(\beta\)-3&  0.35\\
         TGF\(\beta\)-2&  0.84\\
         \hline
    \end{tabular}
    \caption{Resulting \textit{p} values for each cancer vs control cytokine t-test.}
    \label{tab:cytokine_cancer_v_control}
\end{table}

\begin{table}[t]
    \centering
    \begin{tabular}{|c|c|c|c|}
        \hline
         &  IL-8&  TGF\(\beta\)-1&  TGF\(\beta\)-2\\
         \hline
         ESD&  -0.30&  -0.35&  -0.29\\
         EAC&  0.32&  0.24&  0.20\\
         \hline
    \end{tabular}
    \caption{Correlation coefficients between select cytokines and QUS parameters.}
    \label{tab:cytokine_qus_correlation}
\end{table}

\begin{figure*}[!t]
    \centering
    \begin{subfigure}{0.2\linewidth}
        \includegraphics[scale=0.15]{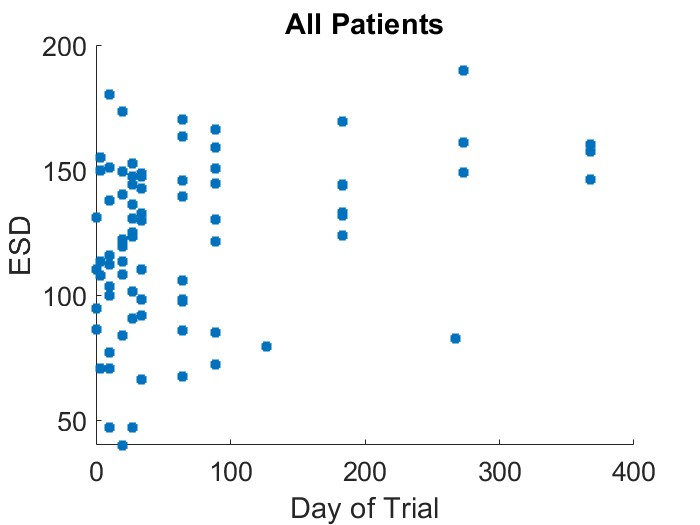}
        \caption{}
        \label{subfig5a:cor_esd}
    \end{subfigure}
    ~
    \begin{subfigure}{0.2\linewidth}
        \includegraphics[scale=0.15]{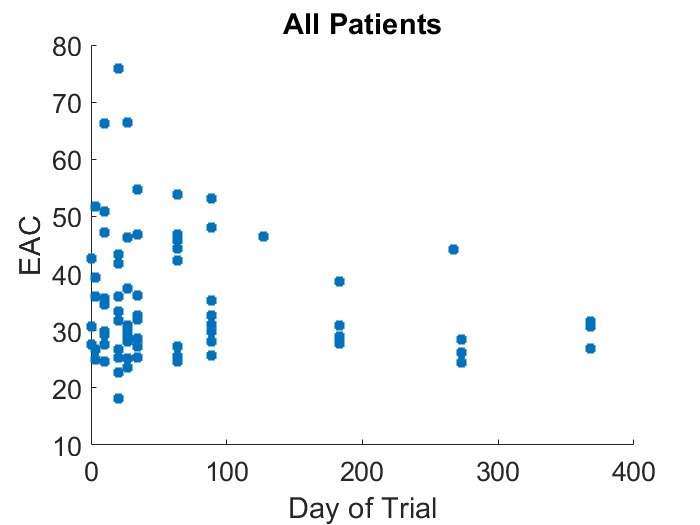}
        \caption{}
        \label{subfig5b:cor_eac}
    \end{subfigure}
    ~
    \begin{subfigure}{0.18\linewidth}
        \includegraphics[scale=0.15]{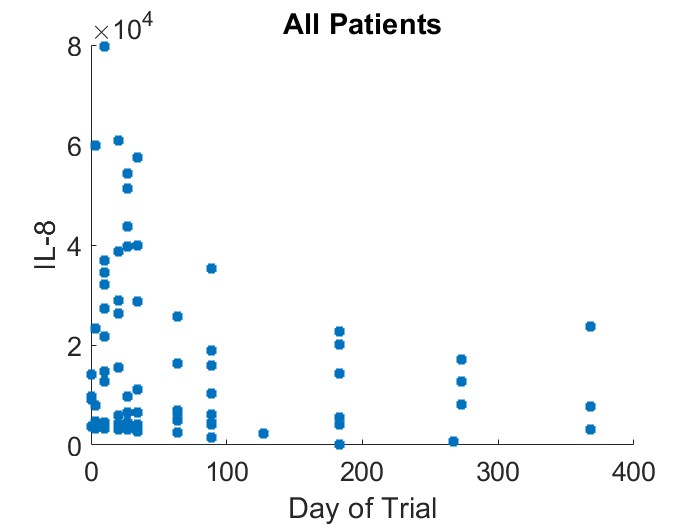}
        \caption{}
        \label{subfig5c:cor_il8}
    \end{subfigure}
    ~
    \begin{subfigure}{0.18\linewidth}
        \includegraphics[scale=0.15]{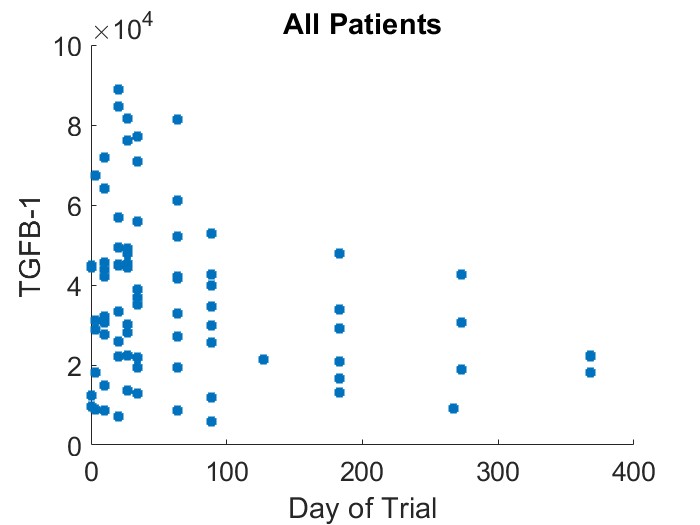}
        \caption{}
        \label{subfig5d:cor_tgfb1}
    \end{subfigure}
    ~
    \begin{subfigure}{0.18\linewidth}
        \includegraphics[scale=0.15]{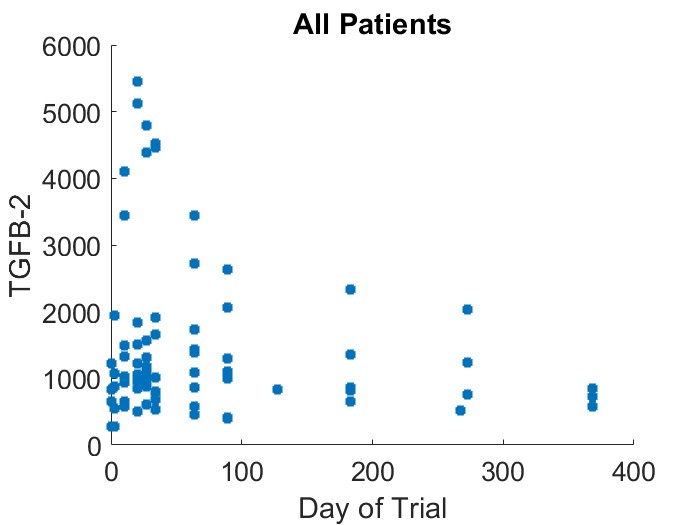}
        \caption{}
        \label{subfig5e:cor_tgfb2}
    \end{subfigure}\\
    \caption{Scatter plots of QUS and cytokine features from all patients that showed clear increases or decreases across time.}
    \label{fig5:qus_cytokine_correlation}
\end{figure*}

\subsection{QUS and Cytokine Correlation}
In these t-tests, MCP-1 resulted in a \textit{p} value of 1.12 x 10-4, which was well below the Bonferroni-corrected significance level of 0.0042 (Table \ref{tab:cytokine_cancer_v_control}). A scatter plot of the MCP-1 data points used for cancer patients versus control patients is shown in Figure \ref{fig4:cytokine_cancer_vs_control}. The protein MCP-1 attracts monocytes to the sites of inflammation caused by injury and infection. Dogs with cancer had much higher levels of MCP-1 than healthy control patients, suggesting that this protein was more active in attracting monocytes to the tumor site in cancer patients. Macrophage phenotype (M1 protumor vs M2 antitumor) was not evaluated but our data suggest this as a future investigation. The other cytokines did not produce significant \textit{p} values in these t-tests (Table \ref{tab:cytokine_cancer_v_control}). 

A few cytokines and QUS parameters showed clear spikes early in the trial, then proceeded to increase or decrease over time as patients went through the trial. Scatter plots of these cytokine data points and QUS data points for all patients are shown in Figure \ref{fig5:qus_cytokine_correlation}. The specific cytokines which showed these trends were IL-8, TGF\(\beta\)-1, and TGF\(\beta\)-2, while the QUS parameters were ESD and EAC. The average magnitude of change, comparing baseline (pre-RT) data with data at day 183 of the trial was +14\% for ESD, -9.2\% for EAC, -52\% for IL-8, -13\% for TGF\(\beta\)-1, and +22\% for TGF\(\beta\)-2. The correlation coefficients for each combination of these cytokines and QUS parameters are given in Table \ref{tab:cytokine_qus_correlation}. Most other correlation coefficients between different cytokines and QUS parameters had values below 0.1, suggesting no correlation. 

The correlation between TGF\(\beta\)-1 and ESD is interesting, as it was the strongest correlation of the dataset. TGF\(\beta\)-1 is a protein that helps control cell growth, cell division, and cell death. It is also a major driver in remodeling the microenvironment during cancer progression, and plays an integral role in post-radiation fibrosis. These results might suggest that in patients with lower amounts of this protein, there was less control over cell growth. Therefore, the ESD, or cell size, in the tumor was larger, creating this negative correlation between ESD and TGF\(\beta\)-1. However, a correlation coefficient of -0.35 only suggests that these features are weakly correlated with each other.

\section{Conclusion}
Quantitative ultrasound is a novel way of characterizing tumor response and detecting LN metastasis. Previous studies have found that QUS parameters can differentiate responders from non-responders due to radiation or chemotherapy \cite{sannachi_non-invasive_2015,osapoetra_assessment_2021}, as well as metastatic versus non-metastatic LNs \cite{mamou_three-dimensional_2010,mamou_three-dimensional_2011}. However, these studies were conducted in well-controlled conditions in preclinical murine research studies of cancer and treatment. Other studies were conducted in human patients with either locally advanced breast cancer or head and neck cancer. In either case, the populations were less diverse compared to the present study with multiple breeds of dogs and multiple different types of cancer. They also did not always take into account inter-patient variability, which makes trends in QUS data harder to truly connect to response or metastasis, as opposed to random biological differences across patients. Since murine preclinical research is conducted in artificial, induced disease states, findings will not necessarily translate to human subjects. The genetic variability, intact immune system, and occurrence of spontaneous cancer in companion animals in our study provided an opportunity to improve the translation of preclinical findings to human studies. Furthermore, our paired study design and statistical tests provided a new perspective on interpreting QUS data which accounts for inter-patient variability. However, our data found that QUS parameters were not distinctly different enough to distinguish between the responders and non-responders, or to distinguish metastatic from non-metastatic lymph nodes, given the small sample size, the heterogeneity of the population, and different types of cancer. Our preliminary data suggests that more subjects should be evaluated to determine how to apply these findings to human studies.

\section*{Declaration of Competing Interest}
The authors declared no potential conflicts of interest with respect to the research, authorship, and/or publication of this article.

\section*{Data Availability}
Data will be made available upon request.

\section*{Acknowledgements}
This work was funded in part by grants from the National Institutes of Health (NIH) (R01CA273700 and R01CA251939), and by the Cancer Center at Illinois (CCIL grant \# 7879). The funding sources had no other involvement in the design of the study, collection or analysis of the data, or writing the paper.

 \bibliographystyle{elsarticle-num-names} 
 \bibliography{vetmed.bib}

\end{document}